\documentclass
[showkeys,showpacs,superscriptaddress,10pt,balancelastpage,nofootinbib,twocolumn,nobibnotes,fleqn]{revtex4}%
\usepackage[verbose,colorlinks,hyperindex,breaklinks=true,pdfusetitle,citecolor=blue]{hyperref}
\usepackage{amsfonts}
\usepackage{amsmath}
\usepackage{amssymb}
\usepackage{graphicx}
\usepackage{color}
\usepackage{epsfig}
\usepackage{dcolumn}
\begin{document}

\title{Non-Dispersive, Accelerated Matter-Waves}
\author{Farhan Saif\footnote{farhan.saif@qau.edu.pk}}
\author{Khalid Naseer}
\author{Muhammad Ayub}
\affiliation{Department of Electronics, Quaid-i-Azam University, Av. III, 45320, Islamabad, Pakistan.}

\begin{abstract}
It is shown that under certain dynamical conditions a material wave packet 
displays coherent, non-dispersive accelerated evolution in gravitational field over a modulated atomic mirror. The phenomenon takes place as a consequence of simultaneous presence of the dynamical localization and the coherent Fermi acceleration for the same modulation amplitude. It is purely a quantum mechanical effect as the windows of modulation strengths supporting dynamical localization and Fermi acceleration overlap for larger effective Plank constant. Present day experimental techniques make it feasible to realize the system in laboratory.
\end{abstract}
\maketitle
\section{Introduction}

Dispersion of a wave packet, with the exception of harmonic oscillator, is a generic phenomena in quantum mechanics. Systems with mixed phase space however display different dynamics near nonlinear resonances and show a local transition of inherently nonlinear spectrum to linear spectrum, resulting a local non-dispersive dynamics. Quantum mechanically, these non-dispersive wave packets are time periodic eigenstates of the Floquet Hamiltonian, localized in the non-linear resonance islands~\cite{SaifPR, buchleitner2002}. In contrast, suppression of dispersion in global dynamics is a manifestation of quantum interference and is an evidence of dynamical localization~\cite{casati1987,Saif1998}, that is regarded as a signature of quantum chaos in a dynamical system. In addition, long time quantum flights due to the accelerator mode islands have been found in dynamical systems~\cite{SaifPRA2007}. In the present paper, we combine the two phenomena and explain the existence of accelerated and dynamically localized wave packet dynamics in Fermi accelerator for certain windows of modulation amplitudes.

In recent years coherent control of matter waves~\cite{BlochRMP2008,Wineland} have introduced enormous applications ranging from quantum metrology \cite{GiovannettiNatPho2011} to quantum corrals \cite{XiongPRA2010}. Introduction of periodic external modulation led to newer avenues to study dynamical systems~\cite{KrivolapovPRA2011,raizen1999} and understand coherent transport \cite{ZangPRA2010,Ben1996,SaifPR}, dynamical localization \cite{saifpla, Milburn}, dynamical revivals~\cite{SaifPRE, ayub2012,fsaifjopt}, dynamical tunneling~\cite{Hanggi,Hanggi1}, coherent acceleration of matter wave in driven systems \cite{Wilkinson,PottingPRA2001} and precision measurement of gravitational acceleration \cite{PoliPRL2011}. Presence of dynamical localization ensures a saturated or very slow dispersive evolution of a quantum particle in a dynamical system, contrary to corresponding classical counterpart that displays global diffusion. Fermi acceleration, another fascinating characteristic of a dynamical system in classical domain~\cite{leonel} defines a steady and unbounded acceleration of a particle as a function of time leading to a non-thermal gain of energy. 

In the present paper, we consider a cloud of ultra-cold atoms propagating along gravitational field and bouncing off a spatially modulated atomic mirror. The system displays dynamical localization~\cite{Saif1998} and Fermi acceleration~\cite{SaifPRA2007} in specific windows of modulation strengths. We show that a parametric control makes it possible to overlap these windows to obtain non-dispersive, coherent acceleration of matter waves. Hence, in the present contribution we report the existence of accelerated dynamics of matter wave superimposed by dynamical localization in Fermi accelerator.

The paper is organized as follows: In Sec.~\ref{model} we present the suggested experimental model. In Sec.~\ref{windows} we briefly discuss conditions for the simultaneous presence of coherent acceleration and dynamical localization in the Fermi accelerator. Later we explain the phenomena in Sec.~\ref{discussion}.

\section{The Experimental Model} 
\label{model}

We consider a cloud of ultra cold atoms that moves along the vertical, $\tilde{z}$-direction under the influence of gravity and bounces off an atomic mirror \cite{Weiss}. The mirror is made up of laser beam incident on a glass prism and undergoing total internal reflection, therefore creating an optical evanescent wave of intensity $\textit{I}(\tilde{z})=\textit{I}_{0}exp(-2k\tilde{z})$. Here, $k^{-1}$ defines characteristic decay length of the field outside of the prism.

We consider that the laser intensity is modulated by an acousto-optic modulator, viz.,
\begin{equation}
I(\tilde{z},\tilde{t})=I_{0} e^{-2\kappa\tilde{z}+\epsilon \sin(\omega \tilde{t%
})},  \label{eq1}
\end{equation}
where, $\omega$ is the frequency and $\epsilon$ is the amplitude of modulation. 
The laser frequency is tuned far from any atomic transition frequency, for the reason the atoms 
effectively stay in their ground states and experience radiation pressure. 
The excited atomic state is adiabatically eliminated and the center-of-mass motion of an atom of 
mass $m$ is governed by one-dimensional effective Hamiltonian, that is,
\begin{equation}
\tilde{H}=\frac{\tilde{p}^{2}}{2m}+ mg\tilde{z} + \frac{\hbar
\Omega_{eff}}{4}e^{-2k\tilde{z}+\epsilon \sin(\omega \tilde{t})}.
\label{eq2}
\end{equation}
Here $\tilde{p}$ is the atomic momentum along the gravitational field and $g$ is the acceleration due to gravity.

We introduce dimensionless position, $z\equiv{\frac{\tilde{z}\omega^{2}}{g}}$, momentum coordinates,
$p\equiv{\frac{\tilde{p}\omega}{mg}}$, and scaled time $t\equiv{\omega\tilde{t}}$. 
In addition, we write $V_{0}\equiv{\frac{\hbar\omega^{2}\Omega_{eff}}{4mg^{2}}}$, 
the inverse of the scaled decay length $\kappa\equiv{\frac{2kg}{\omega^{2}}}$, and the 
scaled modulation strength $\lambda\equiv{\frac{\omega^{2}\epsilon}{2kg}}$. Hence, in dimensionless form the above Hamiltonian reads
\begin{equation}
H
=(\omega^{2}/mg^{2})\tilde{H}
=\frac{p^{2}}{2}+z+V_{0}\; e^{-\kappa(z-\lambda \sin\textit{t})}.
\label{eq3}
\end{equation}
The Hamiltonian given in Eq.~(\ref{eq3}) controls the dynamics of an ensemble of particles in the Fermi accelerator. The dynamics of a classical particle obeys the condition of incompressibility of the flow \cite{Lichtenterg92}, and the phase-space distribution function $P(z,p,t)$ satisfies the Liouville equation, given as
\begin{equation}
\left\lbrace \frac{\partial}{\partial t}+p\frac{\partial}{\partial z}+\dot{p}\frac{\partial}{\partial p}
\right\rbrace P(z,p,t) =0,
\label{eq4}
\end{equation}
where $\dot{p}=-1+\kappa V_{0}\exp[-\kappa(z-\lambda \sin t)]$ is the effective force experienced by the particle.

In the absence of the modulation of the atomic mirror, the atomic dynamics is integrable. For very weak modulations the incommensurate motion closely follows the integrable evolution and remains stable due to the presence of KAM tori~\cite{Lichtenterg92}. However, a non-negligible modulation makes the classical system non-integrable and leads to complex dynamics~\cite{Saif1998}.

In the quantum regime, the atomic evolution is determined by the corresponding Schr$\ddot{o}$dinger equation
\begin{equation}
i k^{\hspace{-2.1mm}-}\frac{\partial \psi}{\partial t}=\left( \frac{p^{2}}{2}+z +V_{0}exp[-\kappa (z-\lambda \sin t)]\right)\psi ,
\label{eq5}
\end{equation}
where $k^{\hspace{-2.1mm}-}\equiv{\hbar \omega^{3}/(mg^{2})}$ is the dimensionless Plank constant, such that, $[z,p]=i(\omega^{3}/mg^{2})\hbar\equiv{ik^{\hspace{-2.1mm}-}}$.
For short decay length, $\kappa^{-1}$, and the matter wave initially far from the mirror surface, we may approximate the optical potential by an infinite potential barrier at the position $z=\lambda \sin \omega t$. In this limit the matter wave bounces off a hard oscillating surface \cite{Saif1998,Milburn}.
The dependence of dynamical localization and accelerated dynamics of matter wave on the modulation strength, $\lambda$, is discussed below.

\section{Dynamical windows} \label{windows}

The quantum dynamics of a material wave packet in atomic Fermi accelerator manifests dynamical localization for a certain range of modulation strength, that defines a localization window \cite{Saif1998}. The lower bound of the window comes from the classical evolution as the onset of the global diffusion takes place above a critical modulation strength, $\lambda_{(cr)}^{(cl)}\cong0.24$  \cite{Lichtenterg92}. 
However, in quantum domain there occurs a phase transition as the spectrum changes from point spectrum to quasi continuum spectrum for a modulation strength, $\lambda_{(cr)}^{(Q)}>\sqrt{k^{\hspace{-2.1mm}-}}/2$~\cite{oliveria}. Beyond the value of the critical strength the matter wave displays dynamical delocalization~\cite{chiri}. Hence, the two conditions together define the localization window \cite{Saif1998}, viz.,
\begin{equation}
 0.24<\lambda<\frac{\sqrt{k^{\hspace{-2.1mm}-}}}{2}.
\label{eq6}
\end{equation}

Fermi acceleration of the particle in the system occurs for a different range of parameters, explained in Ref.~\cite{SaifPRA2007}. The accelerating modes exist in the system for the sinusoidal modulation of the reflecting surface and for modulation strength $ \lambda $ within a windows defined as, 
\begin{equation}
s\pi\leq{\lambda}<\sqrt{1+(s\pi)^{2}}.
\label{eq7}
\end{equation}
where $s$ takes integer and half integer values. Hence, the material wave packet observes accelerated dynamics for a modulation amplitude given in Eq.~(\ref{eq7}) and for the initial distribution originating from the areas of the phase space supporting accelerated dynamics.

\section{Numerical Results} \label{discussion}

Equations~(\ref{eq6}) and (\ref{eq7}) explain that a bouncing quantum particle behaves differently by tuning modulation strength, $\lambda$, away from one window to another. The quantum dynamics of a material wave packet modifies drastically for larger values of the effective Plank constant, $k^{\hspace{-2.1mm}-}$. It is important to note that as the Plank constant increases the dynamical localization windows broadens and an overlap between the two windows defined by Eqs.~(\ref{eq6}) and (\ref{eq7}) takes place.
Hence, the control over the two windows by means of effective Plank constant, $k^{\hspace{-2.1mm}-}$, leads to accelerated and non-dispersive matter wave, that acts as an {\it atomic bullet}. The Eqs.~(\ref{eq6}) and (\ref{eq7}), therefore, define new windows on modulation amplitude which support accelerated but dynamically localized evolution, that is, 
\begin{eqnarray}
s\pi \leq \lambda \leq \frac{\sqrt{k^{\hspace{-2.1mm}-}}}{2}.
\label{al}
\end{eqnarray} 
Here, $\frac{\sqrt{k^{\hspace{-2.1mm}-}}}{2} \leq \sqrt{1+(s\pi)^{2}}$. The lower bound defined by equation (\ref{al}) indicates unbounded acceleration, whereas the upper bound corresponds to the absence of dynamical delocalization. 

As an example, if $k^{\hspace{-2.1mm}-}=12$ the localization window and the first acceleration window, respectively, are $0.24\leq{\lambda}<1.73$, and $1.57\leq{\lambda}<1.86$. An overlap between the two windows occurs for $1.57\leq{\lambda}<1.73$, which leads to an accelerated non-dispersive dynamics of the matter wave. Whereas, for $k^{\hspace{-2.1mm}-}=14$ the first 
acceleration window defined by Eq.(\ref{eq7}) comes within the localization window. For the reason, accelerated non-dispersive dynamics exists when $1.57\leq{\lambda}<1.86$ for a particle originating from the regions of phase space supporting accelerated dynamics.

\begin{figure}[htt]
\centering
\includegraphics[width=0.50\textwidth]{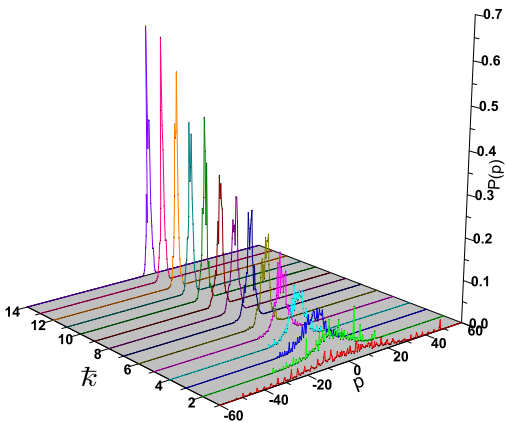} 
\caption{\label{raw} We plot quantum mechanical momentum distributions after a propagation time for $ \lambda=1.7 $ using different scaled Plank constants, $k^{\hspace{-2.1mm}-}$. After a fixed evolution time, $t=1000$, the final momentum distribution exhibit localization with increasing Plank constant. Furthermore, the presence of spikes in the distribution is a consequence of acceleration modes for certain phase space regions.} 
\end{figure}
In order to analyze wave packet evolution, we propagate an atomic wave packet, $\psi(z,0)$, initially in a Gaussian distribution that fulfills the Heisenberg minimum uncertainty criteria and placed above the atomic mirror with initial average momentum $p_{0}=0$. The information about the time evolution of the system is stored in the wave function $\psi(z,t)$ which we obtain by operator splitting technique. We study the probability distributions in position space and in momentum space, which is the marginal integration of the distribution function in phase space, respectively, in momentum and in position space for a fixed time~\cite{2b}.
In Fig.~\ref{raw}, we show the quantum mechanical probability distribution in momentum space, for a propagation time $t=1000$ for different values of $k^{\hspace{-2.1mm}-}$. Here, for different $k^{\hspace{-2.1mm}-}$ the coherent acceleration is prominent as regular spikes in the marginal probability distributions~\cite{SaifPRA2007}. The sharp spikes appear when the modulation strength satisfies the condition, given in Eq.~(\ref{eq7}), and gradually disappear otherwise. These spiky behaviour in momentum distribution is therefore a hallmark of coherent accelerated dynamics. In contrast, the portions of the initial probability distribution originating from the regions of the phase space that do not support accelerated dynamics undergo diffusion and form a background. 
In the momentum distributions the probability confinement in the accelerator islands appears in the form of the spiky nature. As the value of $k^{\hspace{-2.1mm}-}$ is increased the spectrum corresponding to the exponentially localized and accelerated matter wave tends to point spectrum. Hence, the quantum distribution is localized to fewer levels and drops exponentially away from these levels. A careful analysis of the figure~\ref{raw} shows that following Eq.~(\ref{al}) beyond $k^{\hspace{-2.1mm}-}=10$ most of the probability is exponentially confined indicating the localization of the wave packet.

\begin{figure}[htt]
\begin{center}
\includegraphics[width=0.5\textwidth]{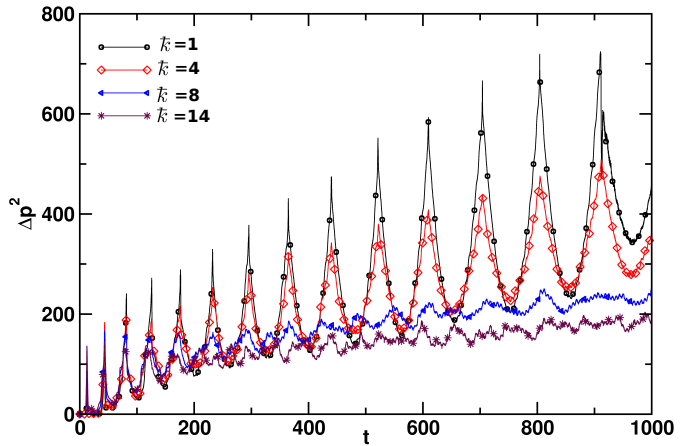} \caption{\label{variance}
We show the variance in the momentum space as a function of time, for a fixed scaled modulation strength $ \lambda=1.7$. Here, we take different values of scaled Plank constant, $k^{\hspace{-2.1mm}-}$. We note that for higher values of $k^{\hspace{-2.1mm}-}$, the acceleration 
window comes within the localization window. As a result the behavior of variance alters and it saturates with evolution time.} 
\end{center}
\end{figure}
In Fig.~\ref{variance} we show the variance in momentum $\Delta p^2\equiv\langle p^2\rangle-\langle p \rangle^2$ as a function of time for different values of $k^{\hspace{-2.1mm}-}$. In all these calculations the modulation strength, $\lambda$ is  fixed to $1.7$. For smaller $k^{\hspace{-2.1mm}-}$ values we observe a linear growth with time due to dynamical delocalization, as discussed in Ref. \cite{SaifPRA2007}. The difference between two peaks in these plots correspond to time of flight of the accelerated wave packet between two impacts with the modulated atomic mirror, the periodicity in these plots indicates coherence in evolution. In addition, gradual increase of the time interval between peaks corresponds to larger time the matter wave takes during its accelerated dynamics between two successive impacts with the modulated surface. Furthermore, drastic variation in $\Delta p^2$ during the time interval reveals different behavior of the wave packet during its evolution over the mirror at the two turning points, one on the surface of the mirror and the other in the gravitational field. For larger values of $k^{\hspace{-2.1mm}-}$, however, dynamical localization restricts the drastic variations resulting a transition to relatively smooth and saturation behavior in $\Delta p^2$ with time. Hence, for larger values of effective Plank's constant, as shown in Fig.~\ref{variance}, we find localized dynamics for the accelerated matter wave.

\begin{figure}
\begin{center}
\includegraphics[width=0.5\textwidth]{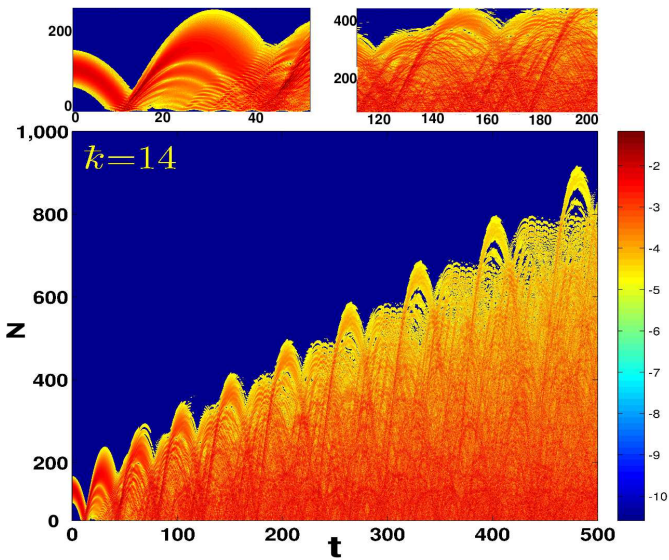} 
\caption{\label{contour1} 
We display contour plots of the quantum mechanical position distribution after a propagation
time, $t=500$, for $\lambda=1.7$ for $k^{\hspace{-2.1mm}-}=14$. } 
\end{center}
\end{figure}

Following classical dynamics in the Fermi accelerator \cite{kn} atomic wave packet follows the classical trajectories and bounces off from the moving surface, as in Fig.~\ref{contour1}. Here, we plot the contour plot of the marginal probability distribution $P(z,t)$ for a propagation time $t=500$ for $\lambda=1.7$ and $k^{\hspace{-2.1mm}-}=14$. Each bounce takes longer time from the previous one, and after every bounce the maximum average position (i.e. the turning point of the wave packet in the gravitational field) becomes higher which corresponds to longer time of flight above the mirror. The two insets show the two turning points, one due to atomic mirror (left) and the other in the gravitation field (right). It is notable that portions of the wave packet following acceleration modes are visible as jet of space-time distributions following classical trajectories. However, the rest of the wave packet reach those areas which do not support coherent acceleration and as a consequence a stochastic background profile appears, shown in the Fig.~\ref{contour1}. 

We study dynamically localized and accelerated dynamics of a material wave packet in the Fermi accelerator by using effective Plank constant, $k^{\hspace{-2.1mm}-}$, as a control parameter. 
The presented theoretical method can be realized in laboratory experiment for cesium atoms of mass $m=2.2\times 10^{-25} \mathrm{kg}$ bouncing off an atomic mirror with a modulation frequency $\omega=7.55 \mathrm{kHz}$ of the external field we find the value of effective Planck's constant $k^{\hspace{-2.1mm}-}=14$. Furthermore, the first localization-acceleration window can be realized in experiment by tuning $\epsilon$ from $0.675$ to $0.80$, provided the decay length of the exponentially decaying field is kept at $k_{L}^{-1}$ $\sim$  $0.8 \mu \mathrm{m}$.


\end{document}